# Second and Third harmonic generation in the opaque region of GaAs

L. Rodriguez[1], J. Trull[1], M. Scalora[2], R. Vilaseca[1], and C. Cojocaru[1]

[1] *Departament de Física, Universitat Politècnica de Catalunya, 08222 Terrassa, Spain*
[2] *Charles M. Bowden Research Center, CCDC AVMC Redstone Arsenal, AL 35898-5000 -U.S.A.*

jose.francisco.trull@upc.edu

**Abstract**

Second and third harmonic generation in the opaque region of a GaAs wafer is experimentally observed both in transmission and reflection. These harmonic components propagate through an opaque material as long as the pump is tuned to a region of transparency or semi-transparency, and correspond to the inhomogeneous solutions of Maxwell's equations with nonlinear polarization sources. We show that measurement of the angular and polarization dependence of the observed harmonic components allows one to infer the different nonlinear mechanisms that trigger these processes, including bulk nonlinearity, magnetic Lorentz and surface contributions. The experimental results are compared with a detailed numerical model that takes into account these different effects.

**Introduction**

The use of semiconductors such as GaAs, GaP or Si in the process of fabrication of nano-devices is at the forefront of modern technology, with the aim of exploiting light-matter interactions at the nanoscale in new and sometimes surprising ways. Light localization at dimensions much smaller than the incident wavelength is possible using the properties of electromagnetic field enhancement by plasmonic resonances at metal-dielectric boundaries, in metamaterials, or under conditions where the dielectric constant of the material approaches zero. At these scales the usual macroscopic theory that describe the behavior of the electromagnetic field should be revisited and analyzed.

Traditionally, the nonlinear (NL) response of these materials has been studied under phase-matching conditions in dielectric materials lacking inversion symmetry, with the principal purpose of transferring energy from the fundamental beam to its harmonics. Under such conditions, the leading NL polarization terms correspond to bulk contributions from the NL potential described through either the quadratic or the third order NL susceptibility tensors. However, other contributions to the NL polarization arise from electric quadrupole-like sources and magnetic interactions and must be considered when the bulk terms either vanish or are drastically reduced, as happens in metals or other materials like GaAs, Si and GaP that may display centrosymmetric behavior as a whole or in particular crystallographic directions. In that case, phase-mismatch plays little or no role.

These circumstances have been considered since the early days of NL optics [1, 2], when different models where advanced to describe surface-generated harmonic signals. It is well-known that under phase mismatched conditions, when a pump pulse crosses an interface between a linear and a NL medium there are always three generated second and third harmonic (SH and TH) components: one is reflected back into the linear medium, and the remaining two are transmitted. This process was first discussed in reference [1], where a mathematical treatment provided a general solution of Maxwell's equation with NL polarization sources, with a homogeneous (HOM) and an inhomogeneous (INHOM) solution. For example, in the absence of absorption, the HOM-SH component travels with the expected group velocity given by material dispersion at the SH wavelength: $k(2\omega)=k_0(2\omega)n(2\omega)$, where $k_0(2\omega)$ is the wavenumber for the SH in vacuum and $n(2\omega)$ is the index of refraction at the SH wavelength. In sharp contrast, the INHOM-SH component is captured by the pump pulse and experiences the same effective dispersion as the fundamental field (FF): $2k_0(\omega)n(\omega)$, where $k_0(\omega)$ is the wavenumber of the FF in vacuum, and $n(\omega)$ is the index of refraction at the pump wavelength. The INHOM signal is referred to as phase locked (PL) component, was experimentally reported for the first time in reference [3], and has been studied in details in reference [4].

A more intriguing situation arises when the fundamental beam is tuned in the transparency range of the material, so that the harmonic wavelengths are tuned well-below the absorption edge. Based on the nature of the INHOM solutions, which travel with the phase and group velocities of the fundamental wave, one may then expect from the Kramer-Kronig relations that the imaginary part of the refractive index experienced by the harmonic components should match that of the fundamental. This conclusion leads to the counter-intuitive hypothesis that the PL harmonic components should be able to propagate inside the material regardless of material dispersion and absorption at the harmonic wavelengths in the opaque region of the spectrum, as long as the pump is at least partially transmitted.

The existence of this phenomenon has been demonstrated experimentally and first reported in reference [5] for a thick GaAs substrate. Subsequently, it was shown that the PL SH component generated in opaque materials can be amplified in a GaAs cavity that displays a resonance only at the FF wavelength [6], and in a GaP substrate where a TH component was generated at 223 nm, in a wavelength range where the dielectric constant is negative [7]. Beyond those contributions, detailed theoretical and experimental studies of NL frequency conversion in the opaque wavelength range are still lacking, most likely because there is widespread perception that the opaque range is intrinsically hostile to light propagation, and thus inaccessible and not very useful. Therefore, in what follows we delve into basic theoretical and experimental investigation of harmonic generation from a GaAs substrate in order to highlight and understand the different mechanisms playing a role in harmonic generation. We then go on to demonstrate that by measuring the polarization and the angular dependence of the generated second and third harmonic signals can lead to detailed information about the relative contributions of each of these mechanisms.

**Model**

We have carried out detailed numerical simulations of these phenomena using the approach outlined in references [8-10]. The method employs a NL Lorentz oscillator model that naturally captures both linear and NL material dispersion, while simultaneously accounting for surface and magnetic nonlinearities. In general, second and third order bulk nonlinearities may be written as $P_i^{NL(2)} = \sum_{j=1,3,} \sum_{k=1,3,} \alpha_{i,j,k} P_j P_k$, and $P_i^{NL(3)} = \sum_{j=1,3,} \sum_{k=1,3,} \sum_{l=1,3,} \beta_{i,j,k,l} P_j P_k P_l$ respectively. Expanding the summations leads to the following second and third order polarizations for isotropic GaAs or GaP having (001) symmetry [10, 11], these expressions generate the following, simplified vector components:

$$\begin{pmatrix} P_x^{NL(2)} \\ P_y^{NL(2)} \\ P_z^{NL(2)} \end{pmatrix} = \begin{pmatrix} (\alpha_{x,y,z} + \alpha_{x,z,y}) P_z P_y \\ (\alpha_{y,x,z} + \alpha_{y,z,x}) P_z P_x \\ (\alpha_{z,x,y} + \alpha_{z,y,x}) P_y P_x \end{pmatrix} = \alpha \begin{pmatrix} P_z P_y \\ P_z P_x \\ P_y P_x \end{pmatrix} \quad (1)$$

and

$$\begin{pmatrix} P_x^{NL(3)} \\ P_y^{NL(3)} \\ P_z^{NL(3)} \end{pmatrix} = \beta \begin{pmatrix} (P_x^2 + P_y^2 + P_z^2) P_x \\ (P_x^2 + P_y^2 + P_z^2) P_y \\ (P_x^2 + P_y^2 + P_z^2) P_z \end{pmatrix} . \quad (2)$$

For simplicity we have consolidated the constants into a single coefficient. For an oscillator system, it can be shown that the constants $\alpha \approx \frac{\omega_0^2}{L}$ and $\beta \approx \frac{\omega_0^2}{L^2}$ [10], where $\omega_0$ and $L$ are the resonance frequency and lattice constant, respectively. In general, NL sources may be specified in terms of a complex dielectric function described by a combined Drude-Lorentz model, which contains a mix of free and bound electrons having one or more resonances. For typical GaAs substrates, free carrier doping (n- or p- type) is rather low, ranging from $10^{14}$ cm$^{-3}$ to $10^{17}$ cm$^{-3}$, so that in the visible and near IR ranges the Drude portion may be neglected. Therefore, in our case only bound electrons are assumed to play a role in surface SH and TH generation. The dielectric functions of GaAs may be described using two Lorentzian functions (see Fig. 1 below), each describing a set of bound charges. Newton's second law for one such species of bound electrons leads to the following effective polarization equation [8-10]:

$$\ddot{\mathbf{P}}_{b1} + \gamma_{b1}\dot{\mathbf{P}}_{b1} + \omega_{01}^2\mathbf{P}_{b1} + \alpha_1\mathbf{P}_{b1}\mathbf{P}_{b1} - \beta_1\left(\mathbf{P}_{b1}\bullet\mathbf{P}_{b1}\right)\mathbf{P}_{b1} = \frac{n_{0b1}e^2}{m_{b1}^*}\mathbf{E} + \frac{e}{m_{b1}^*c}\dot{\mathbf{P}}_{b1}\times\mathbf{H} \qquad (3)$$

where $\mathbf{P}_{b1}$ is the polarization, $\dot{\mathbf{P}}_{b1}$ is the bound current density, $\mathbf{H}$ is the magnetic field, $e$ is the charge of the electron, $m_{b1}^*$ is the bound electron's effective mass, $c$ is the speed of light, and the terms $\alpha_1\mathbf{P}_{b1}\mathbf{P}_{b1}$ and $\beta(\mathbf{P}_{b1}\cdot\mathbf{P}_{b1})\mathbf{P}_{b1}$ are to be interpreted as three-component vectors given by Eqs. (1) and (2). In order to take surface effects into account the field must be expanded near the surface. The development of Eq. (3) is laborious but straightforward [8-10]. For example, an incident field propagating along the z-direction may be written as follows:

$$\mathbf{E} = \begin{pmatrix} E_{\hat{x}} \\ E_{\hat{y}} \\ E_{\hat{z}} \end{pmatrix} = \begin{pmatrix} \hat{\mathbf{x}}\left(E_{TE\hat{x}}^\omega(\mathbf{r},t)e^{i(\mathbf{k}\bullet\mathbf{r}-\omega t)} + E_{TE\hat{x}}^{2\omega}(\mathbf{r},t)e^{2i(\mathbf{k}\bullet\mathbf{r}-\omega t)} + E_{TE\hat{x}}^{3\omega}(\mathbf{r},t)e^{3i(\mathbf{k}\bullet\mathbf{r}-\omega t)} + c.c.\right) \\ \hat{\mathbf{y}}\left(E_{TM\hat{y}}^\omega(\mathbf{r},t)e^{i(\mathbf{k}\bullet\mathbf{r}-\omega t)} + E_{TM\hat{y}}^{2\omega}(\mathbf{r},t)e^{2i(\mathbf{k}\bullet\mathbf{r}-\omega t)} + E_{TM\hat{y}}^{3\omega}(\mathbf{r},t)e^{3i(\mathbf{k}\bullet\mathbf{r}-\omega t)} + c.c.\right) \\ \hat{\mathbf{z}}\left(E_{TM\hat{z}}^\omega(\mathbf{r},t)e^{i(\mathbf{k}\bullet\mathbf{r}-\omega t)} + E_{TM\hat{z}}^{2\omega}(\mathbf{r},t)e^{2i(\mathbf{k}\bullet\mathbf{r}-\omega t)} + E_{TM\hat{z}}^{3\omega}(\mathbf{r},t)e^{3i(\mathbf{k}\bullet\mathbf{r}-\omega t)} + c.c.\right) \end{pmatrix} \qquad (4)$$

with similar expressions for the magnetic fields and polarizations. To be specific, a TM-polarized field will have a single magnetic field polarized along the transverse coordinate x and two electric field components polarized along y and z, respectively. By the same token, a TE-polarized field will have a single electric field component polarized along the x-direction, and two magnetic field components that point along y and z. Then, Eq.(3) may be written as follows:

$$\ddot{\mathbf{P}}_{b1} + \gamma_{b1}\dot{\mathbf{P}}_{b1} + \omega_{01}^2\mathbf{P}_{b1} + \mathbf{P}_{b1}^{(NL)} = \frac{n_{0b1}e^2}{m_{b1}^*}\mathbf{E} + \frac{e}{m_{b1}^*}\left(\mathbf{P}_{b1}\bullet\nabla\right)\mathbf{E} + \frac{e}{m_{b1}^*c}\dot{\mathbf{P}}_{b1}\times\mathbf{H} + ... \qquad (5)$$

where in this case $\mathbf{P}_{b1}^{(NL)}$ contains the second and third order contributions to the total polarization, as given by the expansion of Eqs. (1) and (2) above. We note that all terms in the expansion of Eqs. (1-2) are retained in order to account for the possibility of pump depletion and down-conversion.

Eq. (5) suggests that in the absence of bulk nonlinearities ($\alpha=0$; $\beta=0$; $\Rightarrow \mathbf{P}_{b,2\omega}^{(NL)}=0$), one still has SHG and THG arising from the spatial derivatives of the fields (or polarizations, in the alternative approach presented in [8-10]), which are non-zero at each surface crossing, and the magnetic Lorentz force, which contains intrinsic surface and bulk components since the currents flow inside the bulk and at the surface. Eq. (5) thus describes bound charges that give rise to harmonic generation in dielectrics and semiconductors of arbitrary geometry, as well as in metallic structures, and are valid in the ultrashort pulse regime under conditions of pump depletion. Eq. (5) contains three main contributions that are expected to participate in the resulting NL process: *(i)* the NL bulk contributions arising from the anharmonic spring potentials; *(ii)* a purely surface contribution triggered by the spatial derivatives of the field; and *(iii)* the NL contribution from the magnetic portion of the Lorentz force.

**Numerical simulations**

The complex, local dielectric constant of GaAs [12] is depicted in Fig. 1 (markers). Two resonances can be clearly identified and are fitted using two separate complex Lorentzian functions (solid curves). Eq. (5) describes only one such resonance and is supplemented by another, similar material equation with oscillator parameters that reflect a second resonance. To infer the relative contribution of each one of the terms contained in Eq. (5), we integrate the two oscillator equations together with Maxwell's equations and use the approach outlined in reference [10]. This model solves Maxwell's equations using a spectral method to advance the field, and a predictor-corrector method to solve material equations. We emphasize that by directly integrating a set of NL oscillator equations similar to Eq. (5) one naturally preserves both linear and NL dispersions [10]. In fact, the $\chi^{(2)}$ and $\chi^{(3)}$ are directly proportional to $\alpha$ and $\beta$, respectively, and are never explicitly introduced. All that is required are atomic parameters relating to effective mass, resonance frequency, and approximate atomic diameter or lattice constant.

We consider the propagation of pulses whose temporal duration ranges from 50 fs to 100 fs. Results saturate for longer pulses. A fundamental pulse having a peak power density of ~2 GW/cm$^2$ is tuned to 1064 nm and is incident on a 10 µm-thick GaAs wafer assumed to have (001) crystallographic characteristics, i.e. grown in the direction perpendicular to the surface. Fig. 1 shows the real and imaginary parts of the permittivity of GaAs reported in [12]. One may surmise that the medium is transparent for wavelengths above ~900 nm, strongly absorptive below it, and partially transparent for relatively small thicknesses. Once the dielectric constant is fitted to the two Lorentzian functions, the only free parameters in the equations of motion are the effective mass and the *α* and *β* coefficients. The effective masses of semiconductors can be a small fraction of the electron mass. For GaAs, we are guided by conversion efficiencies, and choose $m^*_{b1} \approx 0.015 m_e$, where $m_e$ is the free electron mass. *α* and *β* are initially chosen using a lattice constant L ~ 3Å and a resonance wavelength between 300 nm and 400 nm. However, as we will see later here too we can adjust these parameters guided by the respective SH and TH conversion efficiencies that we observe. The polarization of the incident fundamental beam is selected to be either TE [ **E** = $(E_{\hat{x}},0,0)$ ], or TM [ **E** = $(0,E_{\hat{y}},E_{\hat{z}})$.] For example, based on Eq. (1), an incident TM-polarized field generates $P_z$ and $P_y$ polarizations that in turn combine to generate a TE-polarized SH signal.

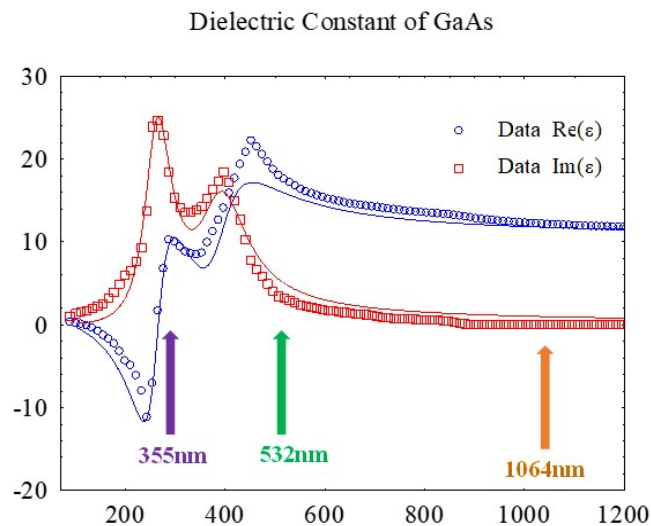

**Figure 1:** Real (blue) and imaginary (red) part of the permittivity.

We have performed simulations where we vary the input beam polarization and the angle of incidence. Figs. 2a - 2c show the transmitted (blue curves, empty circles) and reflected (red curves, empty squares) SH efficiencies as a function of the angle of incidence. As noted earlier, the characteristics of the nonlinearities are such that a TM-polarized pump will generate a TM-polarized SH signal having surface origin – Fig. 2a, while simultaneously the bulk $\chi^{(2)}$ (non-zero α) triggers a TE-polarized SH signal – Fig. 2b. In contrast, if the pump is TE-polarized, only a surface-generated, TM-polarized SH component is triggered – Fig. 2c. The transmitted efficiencies of surface SHG for either TM- or TE-polarized pump are of order $10^{-8}$. Reflection efficiencies are approximately one order of magnitude smaller than transmission, with secondary peaks appearing at large angles. These results indicate different, simultaneously active, interacting polarization channels. The TE-polarized SH vanishes when the pump is TE-polarized. The first important aspects of these results is the appearance of transmitted SH signals, tuned to 532 nm, deep in the opaque region of the material.

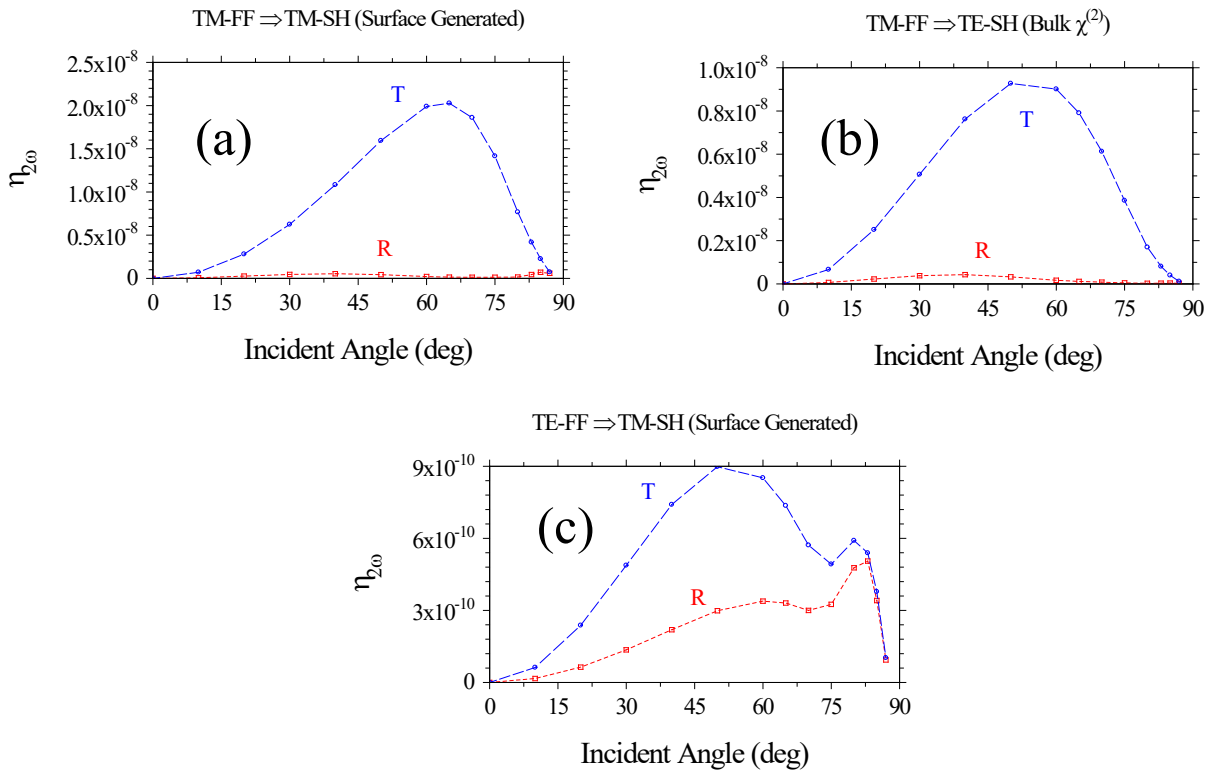

**Figure 2:** Predicted surface-and bulk-generated SH transmission and reflection angular dependence as a function of incident pump polarization. A TM-polarized pump triggers a surface-generated TM-polarized SH signal (a) and a bulk-generated TE-polarized signal, as outlined in the text. A TE-polarized pump (c) is responsible for a TM-polarized SH signal, arising mostly from the magnetic Lorentz nonlinearity.

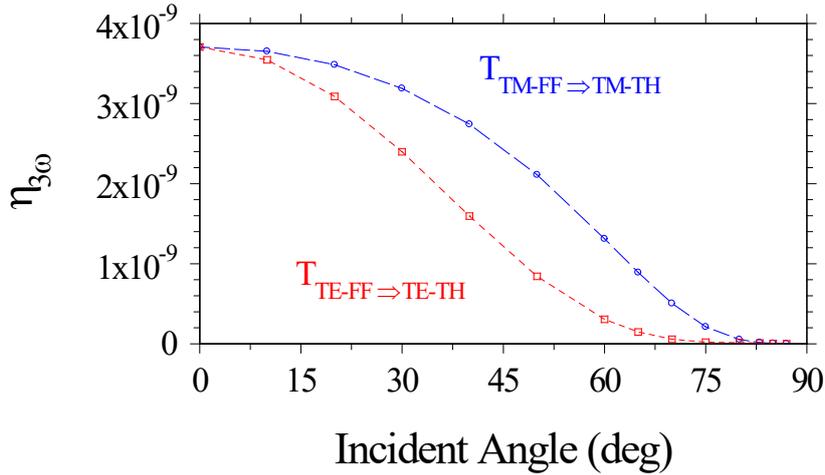

**Figure 3:** Predicted THG for TM- (blue, dashed curve, open circle) and TE-polarized (red, solid curve, open squares) pumps.

THG may be considered in similar fashion. As previously discussed [9], in semiconductors displaying *43m* symmetry the bulk contribution dominates the process. Based on Eq. (2), THG will be triggered with either a TM- or a TE-polarized pump. The particular angular dependence of the transmitted TH signals obtained from the numerical simulations is shown in Fig. 3 for the two possible initial polarization configurations. Reflected THG is more than one order of magnitude smaller in both cases, and follows the angular dependence displayed by the transmitted THG. Quite remarkably, however, the TH is tuned directly to the absorption resonance located near 355 nm, where one would have no expectations to register any transmitted signal.

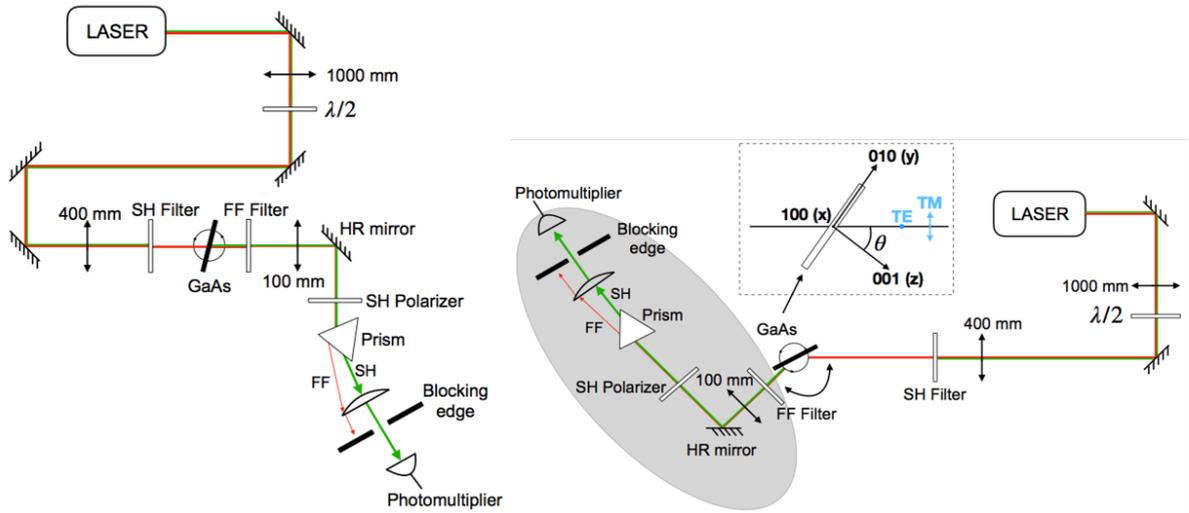

**Figure 4:** Transmission (left) and reflection (right) setups for measuring SH and TH signals.

**Experimental results**

We now describe our experimental observations. The schematic representation of our setup is shown in Fig. 4. The fundamental beam is generated by a fibre laser emitting pulses 13 picoseconds in duration at 1064 nm. Typical power densities used in the experiments were ~ 1.6 GW/cm$^2$. A half-wave plate placed at the laser output controls the polarization of the fundamental beam, allowing selection of either TE or TM polarization. The analysis of the generated SH polarization is performed by a second polarizer placed after the sample. The undoped GaAs wafer 500 μm thick cut with the (001) direction perpendicular to the surface is set on a rotary support that controls the incident angle. The SH at 532 nm is generated at the entrance of the GaAs slab. Since the SH is tuned within the opaque region of the spectrum, only the PL component can propagate and is transmitted through the sample. The inset of Fig. 4 shows the SH components and marks the PL SH component transmitted by the GaAs slab.

The most critical part of the experimental setup is the detection of the generated harmonics, which display efficiencies of the order of 10$^{-7}$ - 10$^{-10}$. In order to detect the faint SH and TH signals, a photomultiplier (Hamamatsu) was used together with a spectral filter having 20 nm band pass transmission band around either the SH or TH frequency. To avoid harmonic generation from other surfaces in the setup, different filters were placed just after the GaAs wafer to attenuate the fundamental beam. Placing the proper filters in the laser path also eliminated possible harmonic signals arising from portions of the setup far from our sample. To separate the weak harmonic radiation from the stronger fundamental beam a prism was used after the GaAs wafer and before the photomultiplier. A stop was placed after the prism to block the fundamental radiation. The entire detection system was mounted on a rigid platform, which was placed on a rotating tail to allow measurement of either transmittance or reflectance. A detailed calibration procedure was performed in order to estimate the efficiencies of the process as accurately as possible. Our setup allows measurement of efficiencies of order $10^{-10}$.

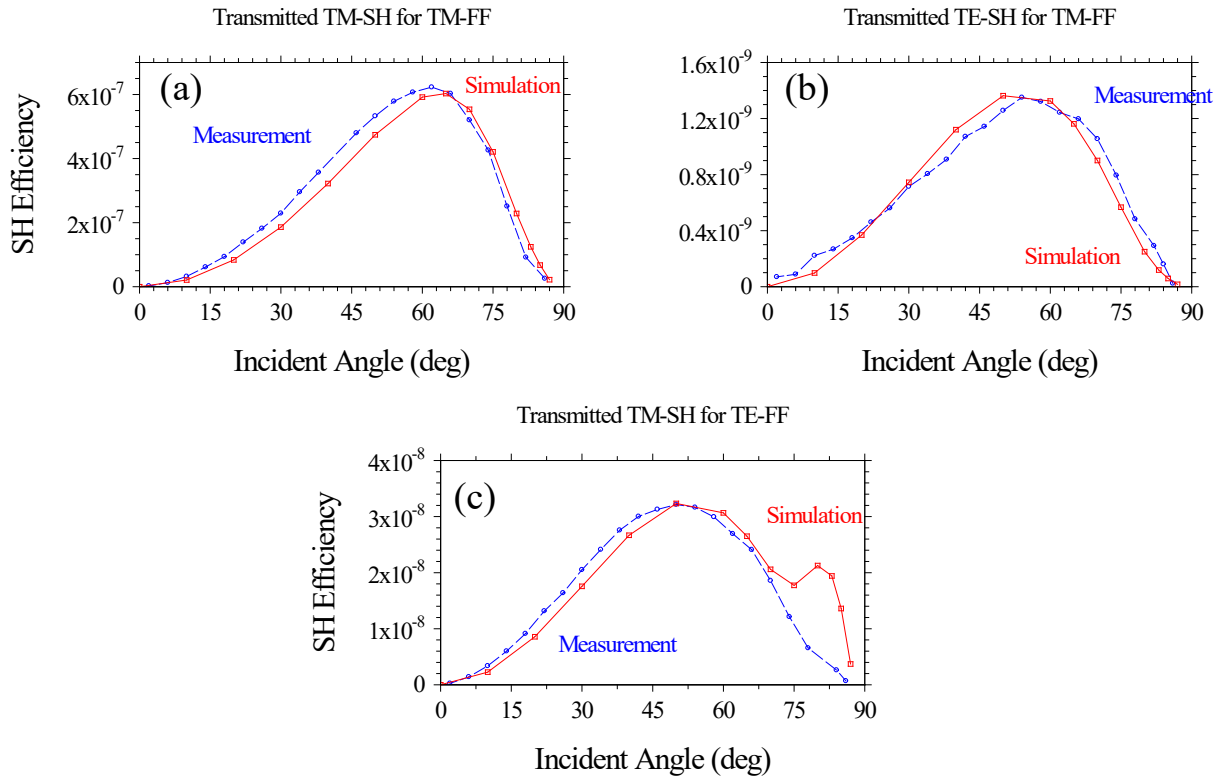

**Figure 5:** Calculated (red, solid curves with open squares) and measured (blue, dashed curves with open circles) SH transmission efficiencies as a function of the angle of incidence and pump polarization, as indicated. The surface generated components [(a) and (c)] agree well with our measurements in both maximum amplitude and shape if the electron's effective mass is chosen to be $m^* = 0.0025 m_e$. The bulk-generated SH efficiency also agrees well with the experiment in both maximum amplitude and shape if the coefficient $\alpha$ corresponds to a $\chi^{(2)} \sim 20-40 \text{pm/V}$. The discrepancies between theory and experiment in (c) at large angles may in part be due to interference effects. The simulations are carried out for a 10μm-thick etalon and pulses a few tens of femtoseconds in duration. The experiment is performed using picosecond pulses and 500μm-thick substrate.

Fig. 5 shows the experimental results (blue, dashed curves with open circles) of the transmitted SH efficiency as a function of the incident angle, for the combinations indicated on top of each plot: TM(FF)-TM(SH) – Fig. 5a; TM(FF)-TE(SH) – Fig. 5b; and TE(FF)-TM(SH) – Fig. 5c. The TE-polarized SH is suppressed when the pump is TE-polarized. As can be immediately inferred, a transmitted component of SH is generated and propagated through a 500 μm – thick sample in the presence of strong absorption, corresponding to the propagation of the PL component. The presence of a TM-polarized SH component is a clear indication of the non-negligible contribution of the surface and Lorentz's terms in this process. In our measurements surface and Lorentz magnetic component efficiencies are of order $10^{-7}$ and $10^{-8}$. In contrast, bulk conversion efficiencies [TM(FF)-TE(SH)] are of order $10^{-9}$. As a result, it is clear that surface and Lorentz terms play a more important role in the generation of SH than the bulk for our GaAs sample. To close this section we briefly mention the discrepancies between theory and experiment in Fig.5c, which depicts the TM-polarized SH signal generated by a TE-polarized pump. We will return to this issue below.

The same measurements were also performed in reflection, and the results are shown in Fig.6. In Fig. 6a we report the reflected, surface-generated SH signal. Both shape and amplitude are remarkably similar, including the large peak near 83°. In Fig. 6b we plot the bulk contribution of SHG when the pump is TM-polarized, and identify two issues. First, the measured SH peak is shifted to larger angles by approximately 10°. Second, the maximum amplitude predicted by the simulation is nearly a factor of 13 smaller than the measured signal. Finally in Fig. 6c we compare the SH efficiency of the TM-polarized beam when the pump is TE-polarized. While both signals are strongly peaked at large angles, the result of the simulation shows a somewhat more complicated structure than the measured signal. As before, the TE-SH signal is suppressed for TE-polarized pump.

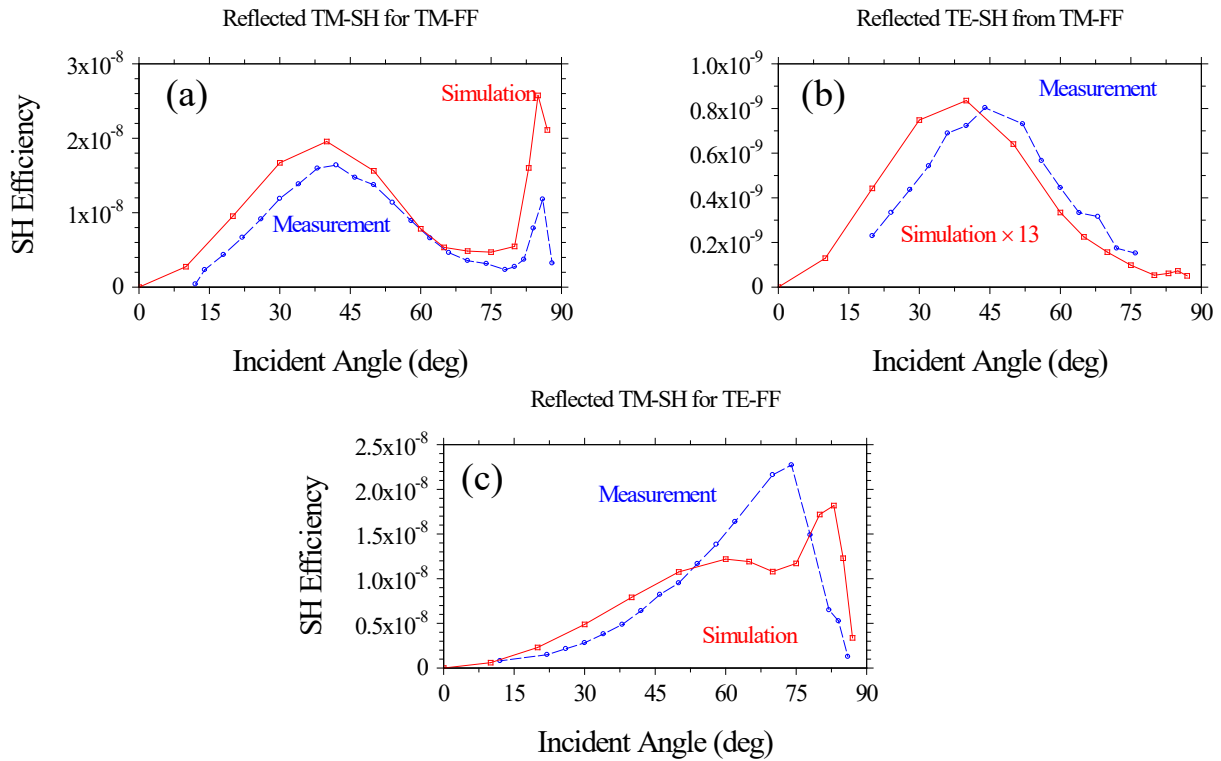

**Figure 6:** Same as Fig.5 but for the reflected components.

The transmitted TH efficiencies were also measured as a function of incident angle and are summarized in Fig.7, where we show measured and simulated transmitted signal for TM- (7a) and TE-polarized (7b) pump pulses. We note that the measured data agrees relatively well with our predictions.

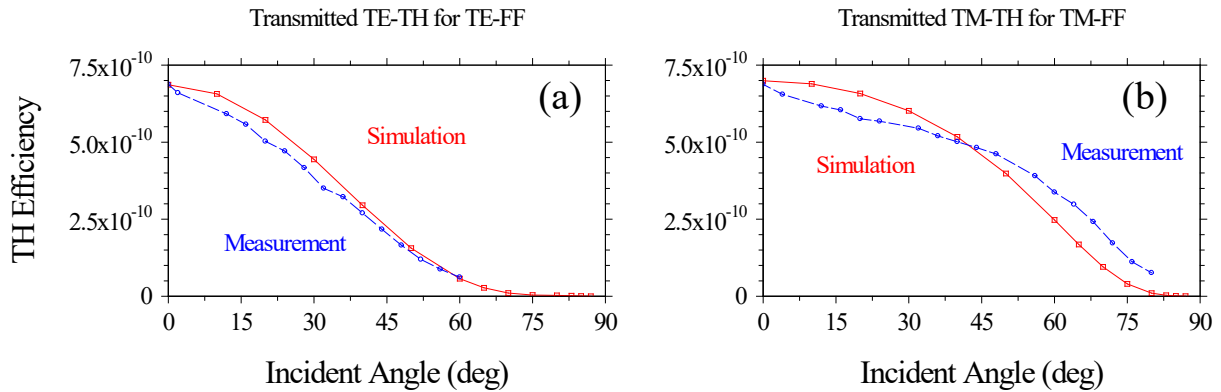

**Figure 7:** Calculated (red, solid curves, open squares) and measured (blue, dashed curves, open circles) for TM-(a) and TE-polarized (b) TH efficiencies as functions of the angle of incidence.

**Discussion and Summary**

As we have seen most of the issues raised by the experiments relate to predictions of the reflected, bulk-generated, TE-polarized SH component when the pump field is TM-polarized, i.e. Fig.6b. In our calculations we have assumed a GaAs substrate of uniform composition from entrance to exit, having no relevant surface features. However, it is known that MBE-grown GaAs may display extended regions of space charges, or depletion layers, due to the existence of surface states that change the symmetry of the bulk inside a thin surface region [13]. In addition, the surface may be either Ga- or As-rich, or have a thin gallium oxide ($Ga_2O_3$) layer that also displays a dielectric anisotropy [14] that may result in direction-dependent effective masses, damping coefficients and resonance frequencies. Notwithstanding the differences that we have highlighted above, we find that the nonlinear oscillator model exemplified by Eq.(7) predicts rather well surface SHG, with most discrepancies arising from bulk-generated SH signal.

In summary, we have experimentally demonstrated SH and TH generation in the opaque region of GaAs, corresponding to phase-locked harmonic generation. We have also shown that the experimental data fits well the numerical simulations performed using a model including different contributions to the nonlinear process arising from bulk and surface terms. The polarization dependence of the measured harmonics allows one to discern the relative contribution of each term to the overall NL conversion efficiency. A detailed study of these phenomena can be used to obtain relevant information about the material parameters, such as the effective mass of bound electrons, second- and third-order NL susceptibilities, and possibly surface states. We were able to measure transmission efficiencies of order $10^{-7}$-$10^{-10}$ for the SH and $10^{-10}$ for the TH. Although propagation phenomena and NL frequency conversion studies below the absorption edge of semiconductors are still lacking primarily because these processes are thought to be uninteresting and inefficient due to absorption and to the naturally high degree of phase mismatch, harmonic generation in absorbing materials can be useful for realizing coherent sources and for the many potential applications that semiconductors find in optical technology.

**References**

1. J. A. Armstrong, N. Bloembergen, J. Ducuing, and P. S. Pershan, *Interactions between Light Waves in a Nonlinear Dielectric,* Phys. Rev. **127**, 1918 (1962).
2. N. Bloembergen, R. K. Chang and C. H. Lee, *Second harmonic generation of light in reflection from media with inversion symmetry*, Phys. Rev. Lett. **16**, 986 (1966).
3. L. D. Noordam, H. J. Bakker, M. P. de Boer, and H. B. van Linden van den Heuvell, *Second-harmonic generation of femtosecond pulses: observation of phase-mismatch effects,* Opt. Lett. **15**, 24 (1990).
4. V. Roppo, M. Centini, C. Sibilia, M. Bertolotti, D. de Ceglia, M. Scalora, N. Akozbek, M. J. Bloemer, J. W. Haus, O. G. Kosareva, and V. P. Kandidov, *Role of phase matching in pulsed second-harmonic generation: Walk-off and phase-locked twin pulses in negative-index media*, Phys. Rev. A **76**, 033829, 2007.
5. Marco Centini, Vito Roppo, Eugenio Fazio, Federico Pettazzi, Concita Sibilia, Joseph W. Haus, John V. Foreman, Neset Akozbek, Mark J. Bloemer, and Michael Scalora, *Inhibition of Linear Absorption in Opaque Materials Using Phase-Locked Harmonic Generation,* Phys. Rev. Lett. **101**, 113905 (2008).
6. V. Roppo, C. Cojocaru, F. Raineri, G. D'Aguanno, J. Trull Y. Halioua, R. Raj, I. Sagnes, R. Vilaseca and M. Scalora, *Field localization and enhancement of phase-locked second- and third-order harmonic generation in absorbing semiconductor cavities*, Phys. Rev. A **80**, 043834 (2009).
7. V. Roppo, J. Foreman, N. Akozbek, M.A. Vincenti, and M. Scalora, *Third harmonic generation at 223 nm in the metallic regime of GaP,* Appl. Phys. Lett. **98**, 111105 (2011).
8. M. Scalora, M.A. Vincenti, D. de Ceglia, V. Roppo, M. Centini, N. Akozbek, and M.J. Bloemer. *Second- and third-harmonic generation in metal-based structures*, Phys. Rev. A **82**, 043828 (2010).
9. M. Scalora, M. Vincenti, D. de Ceglia, N. Akozbek, V. Roppo, M. Bloemer and J. W. Haus, *Dynamical model of harmonic generation in centrosymmetric semiconductors at visible and UV wavelengths*, Phys. Rev. A **85**, 053809 (2012).
10. M. Scalora, M. Vincenti, D. de Ceglia, C.M. Cojocaru, and J. W. Haus, *Nonlinear Duffing oscillator model for third harmonic generation*, JOSA B **32**, 2029-2037 (2015).
11. R. W. Boyd, *Nonlinear Optics*, (Academic, 2003).

12. E. D. Palik, Handbook of Optical Constants of Solids Academic, New York (1985).
13. Jining Qi, "Nonlinear Optical Spectroscopy of Gallium Arsenide Interfaces," University of Pennsylvania, 1995.
14. M. Rebiena, W. Henrion, M. Hong, J. P. Mannaerts, M. Fleischer, *Optical properties of gallium oxide thin films*, Appl. Phys. Lett. **81**, 250 (2002)